\documentclass{emulateapj}

\shorttitle{Spectral properties of 3D-MHD Flows}
\shortauthors{Ohsuga, Kato, and Mineshige}

\newcommand{\lsim}{\raisebox{0.3mm}{\em $\, <$} \hspace{-2.8mm}
\raisebox{-1.8mm}{\em $\sim \,$}}
\newcommand{\gsim}{\raisebox{0.3mm}{\em $\, >$} \hspace{-2.8mm}
\raisebox{-1.8mm}{\em $\sim \,$}}

\begin{document}

\title{Spectral properties of 
Three-dimensional Magneto-hydrodynamical Accretion Flows}


\author{K. Ohsuga\altaffilmark{1,2}, 
Y. Kato\altaffilmark{2}, and S. Mineshige\altaffilmark{2}}


\altaffiltext{1}{Department of Physics, Rikkyo University, 
Toshimaku, Tokyo 171-8501, Japan}
\altaffiltext{2}{Yukawa Institute for Theoretical Physics, Kyoto University,
Kyoto 606-8502, Japan}


\begin{abstract}
In spite of a large number of global three-dimensional (3D) 
magneto-hydrodynamical (MHD) simulations of accretion flows and jets
being made recently, their astrophysical relevance for realistic
situations is not well known.
In order to examine to what extent the simulated MHD flows can 
account for the observed spectral energy distribution (SED) of 
Sagittarius A* (Sgr A*), for the first time we calculate 
the emergent spectra from 3D MHD flows in a wide range of wavelengths
(from radio to X-ray) by solving the 3D radiative transfer equations.
We use the simulation data by Kato, Mineshige, \& Shibata (2004) and
perform Monte Carlo radiative transfer simulations,
in which synchrotron emission/absorption, free-free
emission/absorption, and Compton/inverse Compton scattering are taken
into account.  We assume two temperature plasmas and calculate
electron temperatures by solving the electron energy equation.
Only thermal electrons are considered.  

It is found that the 3D MHD flow generally over-produces X-rays by
means of bremsstrahlung radiation from the regions at large radii.
A flatter density profile, 
$\rho \propto r^{-a}$ with $a<1$,
than that of the advection-dominated accretion flow (ADAF), 
$\rho\propto r^{-3/2}$, 
is the main reason for this.
If we restrict the size of the emission region to be as small as
$\sim 10~r_{\rm s}$, where $r_{\rm s}$ is the Schwarzschild radius,
the MHD model can reproduce the basic features of the observed
SED of Sgr A* during its flaring state.  Yet,
the spectrum in the quiescent state remains to be understood.  We
discuss how to resolve this issue in the context of MHD flow models.
Possibilities include modifications of the MHD flow structure
either by the inclusion of radiative cooling and/or significant
contributions by nonthermal electrons.  It is also possible that the
present spectral results may be influenced by particular initial
conditions.
We also calculate the time-dependent spectral changes, finding that
the fluxes fluctuate in a wide range of the frequency and the flux at
each wavelength does not always vary coherently.
\end{abstract}


\section{INTRODUCTION}
There is a long research history in the theoretical
modeling of black-hole accretion flows.
The standard-disk picture was first established by Shakura \& Sunyaev
(1973) after many attempts.  In this model the gravitational energy is
efficiently converted to radiation energy and is finally radiated
away.  Then, the disk was predicted to be luminous and relatively
cold, exhibiting multi-color blackbody spectra (Mitsuda et al. 1984).
The standard-disk model is widely accepted as a model for disks with
moderately high accretion rates, 
$\dot{M} \lsim L_{\rm E}/c^2$ 
with $L_{\rm E}$ being the Eddington luminosity and
$c$ being the light velocity (Esin, McClintock, \& Narayan 1997).  
In fact, the calculated spectra based on the standard disk model
fits well with the thermal component of the high-state of the black
hole candidates (BHCs, Ebisawa 1999), including the specific
temperature gradient (Mineshige et al. 1994), and probably the big
blue bump of the active galactic nuclei (AGNs, Shields 1978; Malkan
1983).

On the other hand, when the mass-accretion rate is much less than the
critical value of $\sim L_{\rm E}/c^2$, the radiation loss is
inefficient and thus the accretion flow becomes a radiatively
inefficient accretion flow (RIAF).  In such a situation, the thermal
energy of the gas can be advected inward towards the black hole
without much being radiated away.
This basic idea, now known as the notion of advection-dominated
accretion flow (ADAF), was first presented by Ichimaru (1977)
and has been investigated in quite a lot of detail since the 1990's
(Narayan \& Yi 1994, 1995a, b; Abramowicz et al. 1995; 
for reviews, see Narayan, Mahadevan, \& Quataert 1998; 
Kato, Fukue, \& Mineshige 1998).
Since the ADAF can reproduce hard, power-law spectra in the X-ray range,
as has been observed, this model was thought to be a good
representation of disks in low-luminosity AGNs (LLAGNs) 
and in the BHCs during their low-hard state.
Remarkably, the ADAF model can nicely fit the 
emergent spectrum of Sagittarius A* (Sgr A*) and other black-hole
objects
(Narayan, Yi, \& Mahadevan 1995; Manmoto, Mineshige, \& Kusunose 1997;
Narayan et al. 1998; Manmoto 2000; 
\"Ozel, Psaltis, \& Narayan 2000; Oka \& Manmoto 2003).
(We should keep in mind, however, that most of observational data of
Sgr A* only gives upper limits except at radio and X-ray wavelengths.)

We should be aware that there are a number of serious problems
inherent to the ADAF formulation.  For example, the ADAF model cannot
properly treat three-dimensional motion, because the ADAF model is a
one-dimensional model (in the sense that only the radial structure is
solved), although both of the simple arguments and hydrodynamical
simulations have shown that the occurrence of convections and/or
outflow seems to be a natural consequence in the RIAF (Narayan \& Yi
1994; Igumenshchev \& Abramowicz 2000).  The magnetic fields are
treated as a single parameter, the plasma-$\beta$, to predict
synchrotron emissivity, which is too simple and problematic to
describe the dynamics of magnetic fields (see a comprehensive
discussion in Narayan 2002).  Yuan, Quataert, \& Narayan (2003)
recently succeeded in reproducing the spectrum of Sgr A* by
introducing nonthermal electron components, but they still treat the
magnetic fields in a simplified way.  Recent MHD simulations have
commonly shown the important role of the magnetic fields in the flow
dynamics.  The MHD flow is intrinsically time varying and exhibits
fractal structure (e.g., Kawaguchi et al. 2000; Machida \& Matsumoto
2003, hereafter refereed to as MM03).  To summarize, the lack of
three-dimensional (3D) motion and the over-simplified treatment of
magnetic fields lead to self-inconsistencies in the ADAF solution.

The global 3D MHD simulations of RIAFs were first made by Matsumoto (1999),
and have been extensively performed recently by several groups 
(see Mineshige \& Makishima 2004 for a compilation of recent works).
It has been revealed by the 3D MHD simulations 
that the flow pattern is considerably complicated 
and differs significantly from that of the ADAF model
(Hawley 2000; Machida, Hayashi, \& Matsumoto 2000; Hawley \& Krolik 2001; 
Machida, Matsumoto \& Mineshige 2001, hereafter refereed to as MMM01;
Hawley \& Balbus 2002, hereafter HB02; MM03;
Igumenshchev, Narayan, \& Abramowicz 2003, hereafter INA03).
Stone \& Pringle (2001, hereafter SP01) established that a complex
flow pattern is produced due to a magneto-rotational instability (MRI,
see also Hawley, Balbus, \& Stone 2001; Hawley 2001; Balbus 2003 for a
review).  Since the dynamics of magnetic fields is fully solved, the
3D MHD models seem to be more suitable for describing the RIAF than
other models without appropriate treatment of magnetic fields.  Thus,
the MHD models are expected to fit the observations.  Surprisingly,
however, such a critical test has not been well investigated.
Some papers discussing the observational appearance of the simulated
MHD flow, HB02 for example, have shown the distribution of the
synchrotron emissivity based on their MHD model.  Mineshige et
al. (2002) reported a preliminary examination, claiming that the
radial density profile is significantly flatter than that of ADAF,
leading to over-production of X-rays, but they calculated the spectra
in a simplified way.  Their results should be confirmed by full 3D
radiative transfer calculations.  More recently, Goldston, Quataert,
and Igumenshchev (2005) calculated the spectra based on the simulated
MHD flow by INA03, but only in the radio band.  To our best knowledge
nobody has yet calculated the emergent spectra in a wide range of
wavelengths (from radio to X-ray).  This prompted us to calculate, for
the first time, the emergent spectra based on simulated 3D MHD flows
by performing 3D Monte Carlo radiative transfer simulations and
directly comparing them with the observed data of Sgr A*. 

The plan of this paper is as follows:
We present our model and the method of Monte Carlo radiative transfer 
in \S 2.  The results will be displayed with the observational data
in \S 3.  We will demonstrate a serious discrepancy between them.  A
discussion of the loopholes, which cause the discrepancies, and how to
remove them will be given in \S 4.  The final section is devoted to
conclusions.

\section{MHD MODEL AND METHOD OF RADIATIVE TRANSFER CALCULATIONS}
\subsection{Overview of Adopted MHD simulations}
Our spectral calculations are based on 
3D MHD simulations by Kato, Mineshige, \& Shibata 
(2004: hereafter refereed to as KMS04).
They investigated the evolution of a torus threaded by 
weak localized poloidal magnetic fields.
The Overall evolution of 3D MHD accretion flows 
is divided into two distinct phases (see Figure 1 in KMS04).
In the first phase ($\tilde t\lsim 1800$),
toroidal magnetic fields are generated by differential rotation
and are accumulated in the central region,
driving a magnetic tower jet,
where $\tilde t$ is the time normalized by 
$~r_{\rm s}/c\sim 10^{-5} (M/M_\odot) $s
with $M$ being the black-hole mass
and $r_{\rm s}$ being the Schwarzschild radius. 
The jet is, however, a transient phenomenon. 
Eventually, ($\tilde t> 1800$), the jet ceases 
and a quasi-steady, geometrically thick density distribution 
with complex field configuration is produced.

In the quasi-steady state,
the radial density profile is $\rho\propto r$ in the inner part
($r<20r_{\rm s}$),
while $\rho\propto r^{-1}$ in the outer part ($r>20r_{\rm s}$)
(see Figure 4 in KMS04).  That is, there is a broad peak
at $r \sim 20 r_{\rm s}$ in the density profile,
which might be a remnant of the initial torus (discussed later).
Note that this density profile differs from the results of some 
other 3D MHD simulations (see \S 4 for discussion).
Also note that the density distribution of the ADAF model 
is significantly steeper; $\rho\propto r^{-a}$ with $a=1-3/2$.

Throughout the present study, we set
the black-hole mass to be $M = 2.6 \times 10^6 M_\odot$
(Sch\"odel et al. 2002; Ghez et al. 2003).  Thus,
the normalized time corresponds to $t/\tilde t = $26 s.
We basically use the data of 
quasi-steady state at $\tilde{t}=2210$ ($t=5.7\times 10^4$s), 
when no obvious outflow is observed (see Figure 1 in KMS04),
except at \S 3.3, where we will examine spectral variations.

\subsection{Calculations of Physical Quantities}
In the present study,
we use Cartesian coordinates, ($x,y,z$),
where the black hole is located at the origin of the coordinate axes,
the $z$-axis is set to be the rotation axis of the accretion flow,
and the $x$-$y$ plane corresponds to the equatorial plane.
We employ Cartesian grids with numbers 
($N_x$, $N_y$, $N_z$)=($100,100,197$) of cells.
We assume that radiation is generated within the cylindrical region
with radius, $R$; $(x^2+y^2)^{1/2}\leq R$ and $z\leq 100~r_{\rm s}$.
We consider the cases of $R=10~r_{\rm s}$ and $30~r_{\rm s}$.
The size of the calculating box is $2X\times 2Y\times 2Z$,
where we set $(X, Y, Z)=(R,R,100~r_{\rm s})$.

We take 3D data of density, magnetic fields, and 
proton temperature distributions in the accretion flows
from the 3D MHD simulations (KMS04).
Since the MHD simulation only give the normalized density, $\tilde{\rho}$, 
with the normalization, $\rho_0$,
and the normalized field strength, $\tilde{B}$, 
we have one free parameter, $\rho_0$, to determine absolute values 
of density and magnetic fields; that is,
$\rho = \rho_0 \tilde{\rho}$ and $B = (\rho_0 c^2)^{1/2} \tilde{B}$,
where $B$ is the field strength.
The proton temperature does not depend on the density parameter
and is given by the MHD simulation as
$(\mu m_{\rm p}c^2/k)\tilde{c_s}^2$,
where
$\mu$ is the mean molecular weight (=0.5),
$m_{\rm p}$ is the proton mass,
$k$ is the Boltzmann constant,
and $\tilde{c_s}$ is the normalized sound velocity
obtained by the simulation.

Although KMS04 assumed a one-temperature plasma, 
here we adopt two-temperature assumptions.
Assuming that the electrons have a Maxwellian distribution,
we evaluate the electron temperature, $T_{\rm e}$, 
through the energy balance of the electrons between 
Coulomb collisions with ions and radiative cooling,
\begin{equation}
   \int_{-Z}^{Z} \int_0^{R} 
   \lambda_{\rm ie} 2\pi r dr dz = \int L_\nu d\nu.
\label{ebala}   
\end{equation}
Here, 
$\lambda_{\rm ie}$ is the energy transfer rate from
ions to electrons (Stepney \& Guilbert 1983),
$L_\nu$ is the luminosity at frequency, $\nu$, respectively.
In the present study, 
we suppose, for simplicity, 
the electron temperature to be a function of only the radius, $r$,
and to be independent of the altitude, $z$.
This simplification does not affect the results too much, 
since the emission from regions around the equatorial plane, 
where the density is at a maximum, is dominant at each radius.
In addition, we solve for the electron temperature by dividing
the calculating box into two parts;
the inner ($r<10|r_{\rm s}$) and outer parts ($r=10-30~r_{\rm s}$).
The fractional luminosity,
 $L_\nu$, is obtained by Monte Carlo simulations
(see next subsection) for a given $T_{\rm e}$,
which does not always satisfy equation (\ref{ebala}).
We thus iteratively calculate the electron temperature 
and the emergent spectrum so as to meet the condition (\ref{ebala}).

\subsection{Monte Carlo Radiative Transfer Calculations}

The method of the Monte Carlo simulation is 
based on Pozdnyakov, Sobol, \& Sunyaev (1977). 
In this study, synchrotron emission/absorption, free-free emission/absorption, 
and Compton/inverse Compton scattering are taken into account.
Some of the photons emitted by synchrotron 
and free-free emission pass through the calculating box
without being scattered or absorbed, 
while some are scattered and/or
absorbed in the accretion flow,
depending on the mean free path at the photons emitting position.
In order to efficiently calculate the emergent spectra,
we introduce a weight, $w$, as described by 
Pozdnyakov, Sobol, \& Sunyaev (1977). 
At the beginning,
the weight, $w_0$, is set equal to unity for each emitted photon 
and then we calculate the escape probability, $P_0$.
The escape probability of a photon 
after the $i$-th scattering (for $i\ge 1$), $P_i$, is evaluated as
\begin{eqnarray}
  P_i= \exp \left[
    -\left\{ 
      \left( \frac{\rho(x_i,y_i,z_i)}{m_{\rm p}} \right) 
      \sigma_{\rm KN}(x_i,y_i,z_i) 
      \right. \right. \nonumber\\
    +\kappa^{\rm abs}_\nu(x_i,y_i,z_i)
    \bigl. \bigl. \biggr\} l \biggr]&& ,
\end{eqnarray}
where 
$(x_0,y_0,z_0)$ corresponds to the point
where a photon was generated,
$(x_i,y_i,z_i)$ is the point 
where a photon is subject to the $i$-th scattering,
$m_{\rm p}$ is the proton mass, 
$\sigma_{\rm KN}$ is the Klein-Nishina cross section 
(Rybicki \& Lightman 1979),
$\kappa_{\rm abs}$ is the absorption coefficient,
and $l$ is the distance from the point $(x_i,y_i,z_i)$ 
to the boundary of the calculating box, respectively.
Assuming local thermodynamic equilibrium (LTE),
we can write the absorption coefficient as 
\begin{equation}
\kappa^{\rm abs}_\nu = \frac{\left( \varepsilon^{\rm syn}_\nu
+ \varepsilon^{\rm ff}_\nu \right )}{4\pi B_\nu},
\end{equation}
where $\varepsilon^{\rm syn}_\nu$ and $\varepsilon^{\rm ff}_\nu$ are
the synchrotron and free-free emissivity, respectively
(Pacholczyk 1970; Stepney \& Guilbert 1983)
and $B_\nu$ is the Planck function.
The quantity of $w_0 P_0$ represents the transmitted portion of photons
and is recorded to calculate the penetrated spectrum 
or reprocessed photons according to the escape direction of the photon. 
A fraction of $A_0$ of the remaining portion, $w_1 \equiv w_0 (1 - P_0) A_0$, 
undergoes at least one scattering, 
while $w_0 (1-P_0) (1-A_0)$ describes the absorbed portion,
where $A_i$ ($i\ge 0$) is the scattering albedo.
The transmitted portion of photons after the $i$-th 
scattering, $w_i P_i$, is recorded to calculate the transmitted spectrum,
and the remaining portion, $w_i (1 - P_i) A_i$, undergoes the $(i+1)$-th 
scattering. 
This calculation is continued 
until the weight $w_i$ becomes sufficiently small ($w_i \ll 1$). 
The whole process is simulated by the Monte Carlo method. 
Finally, we suppose the region within the Schwarzschild radius,
$(x^2+y^2+z^2)^{1/2}<r_{\rm s}$,
to be vacuum.
Gravitational lensing as well as photon redshifts are not considered,
since the black hole is much smaller than the 
calculating box in our study.

\section{RESULTS: SPECTRAL FEATURES OF MHD FLOW}
\subsection{Basic features}

We first display the representative spectra of the MHD flow
in Figure 1 (see thick solid curves)
together with the observed data of Sgr A*.
\begin{figure}[b]
\epsscale{1.18}
\plotone{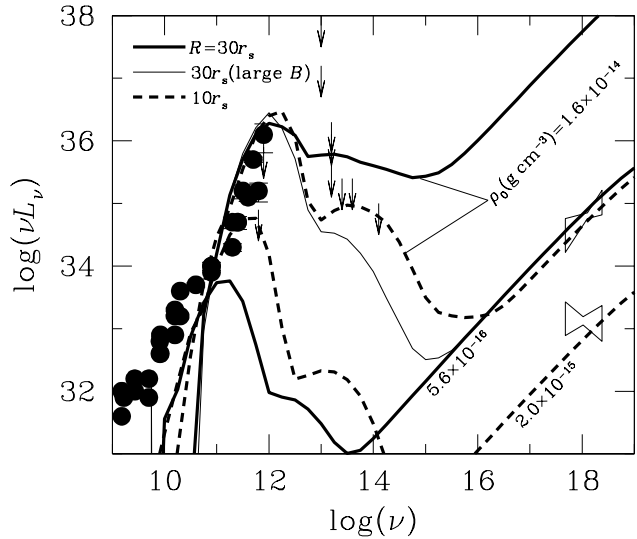}
\caption{
The emergent spectra of the MHD accretion flow
and the observed data of Sgr A*.
The thick solid curves are the resultant spectra for $R=30~r_{\rm s}$.
Here, the adopted density parameters are
$\rho_0=1.6 \times 10^{-14}{\rm g \, cm^{-3}}$ (upper) and
$5.6 \times 10^{-16}{\rm g \, cm^{-3}}$ (lower), respectively. 
The thin solid curve indicates the spectrum, 
where the magnetic fields are set to be 
ten times stronger than the spectrum represented by 
the lower thick-solid curve.
The spectra from the innermost part within 10 $r_{\rm s}$
are plotted by the dashed curves
[$\rho_0=1.6 \times 10^{-14}{\rm g \, cm^{-3}}$ (upper dashed),
$2.0 \times 10^{-15}{\rm g \, cm^{-3}}$ (lower dashed)].
The filled circles and the lines with arrows indicate 
the data and upper limits by radio and IR observations.
For more detailed information regarding the data, please
refer to Narayan et al. (1998).
X-ray observations of the flaring and the quiescent state are 
shown by the two `bowties' (Baganoff et al. 2001, 2003).
\label{fig1}}
\end{figure}
The adopted parameters are the radius of the emitting region, $R=30~r_{\rm s}$,
the elapsed time at $t = 5.7\times 10^4$s, and the density normalization,
$\rho_0=1.6\times 10^{-14}\rm g \, cm^{-3}$ (upper) 
and $5.6\times 10^{-16}\rm g \, cm^{-3}$ (lower).
It might be noted that the emission is predominantly 
 from the equatorial plane and other parts at large altitudes 
($|z|\gg 10~r_{\rm s}$) do not contribute very much to the emergent spectra,
although we sum up all the contributions from the vertically elongated 
cylinder between $Z=\pm 100~r_{\rm s}$.
This is because the density rapidly decreases as $|z|$ increases.
The calculated electron temperatures are listed in Table \ref{Te}.
As expected, electron temperatures are insensitive to radius,
but there is a slight tendency that $T_{\rm e}$ decreases with
an increase of radius.
Similarly, relatively flat electron temperature profiles
were obtained in two-temperature, hot accretion flows 
(Narayan \& Yi 1995b; Nakamura et al. 1996;
Manmoto, Mineshige, \& Kusunose 1997) and evaporated disk-corona flows
(Liu et al. 2002).

\begin{table}[t]
\begin{center}
\caption{Calculated electron temperature.\label{Te}}
\begin{tabular}{ccc}
\tableline\tableline
$\rho_0$ & $T_{\rm e}(r\leq 10~r_{\rm s})$ 
& $T_{\rm e}(10~r_{\rm s}<r<30~r_{\rm s})$ \\
\tableline
{$1.6\times 10^{-14} \rm g\,cm^{-3}$} & {$5.56\times 10^9 \rm K$}
& {$3.45\times 10^9 \rm K$} \\ 
{$5.6\times 10^{-16} \rm g\,cm^{-3}$} & {$4.93\times 10^9 \rm K$}
& {$3.58\times 10^9 \rm K$} \\ 
{$2.0\times 10^{-15} \rm g\,cm^{-3}$} & {$5.20\times 10^9 \rm K$}
& --- \\ 
{$3.0\times 10^{-15} \rm g\,cm^{-3}$}(Our 2D) & {$4.21\times 10^9 \rm K$}
& {$3.65\times 10^9 \rm K$} \\ 
\tableline
\end{tabular}
\tablenotetext{}{
Our 2D: our 2D MHD simulations
}
\end{center}
\end{table}

It is apparent that the resultant spectra look similar to those of the
ADAF model.  The lower-energy peak in the radio band is a synchrotron
peak created because of significant self absorption of 
synchrotron emission.
The IR emission at around $\log\nu=14$ is due to inverse Compton
scattering of synchrotron photons.
Since we assume thermal electrons only, 
the spectral slope at $\log\nu<11$ is $L_\nu \propto \nu^2$,
corresponding to that of Rayleigh-Jeans emission.
The electron temperatures do not vary much, even if we change
$\rho_0$, and, hence, the radio parts (which depends primarily on 
the electron temperature) are roughly identical among different models,
as shown in Figure 1.  In contrast, the flux at $\log\nu > 12$ 
and the frequency of the lower-energy peak (in the radio band)
are both sensitive to $\rho_0$.

Here, we stress that the magnetic field strength affects the 
lower-energy peak in the radio band and the IR flux, whereas 
the X-ray flux and the slope at $\log\nu<11$ do not depend on 
the magnetic field strength itself. To check what alters if
the magnetic field strengths were under or over-estimated in the 
simulation, we calculate the spectra with artificially strengthened
magnetic fields, keeping the same density profile.
The thin solid curve in Figure 1 indicates the spectrum,
here the magnetic field strength is set to be ten times 
stronger than the original values, so as to be compared with the
original (lower thick-solid curve). 
It is found that the X-ray flux and the slope at 
$\log\nu<11$ do not change, but the radio peak luminosity and 
IR flux both increases. This is because the synchrotron emission 
contributes to the radio flux and inverse Compton scattering 
of the synchrotron photons is dominant in the IR band.

A big distinction between the spectra of ADAF and those of
simulated MHD flows is found in the X-ray bands; namely,
inverse Compton scattering is dominated in the former,
while the thermal bremsstrahlung is the dominant mechanism in the
latter.  This reflects a flatter density profile in the MHD flow
than for the ADAF, since a flatter density profile means that more
material is present at large radii than the for case with a steep
density profile, thereby producing more bremsstrahlung photons
(discussed later).

We confirmed that the resultant spectra do not significantly 
change even if we employ the density and magnetic fields 
averaged over the azimuth in the simulated  data of KMS04. 
This implies that 3D effects are not essential for the study 
of the spectra. This is because the flow pattern is almost 
axisymmetric in the simulations of KMS04, who calculated the
evolution of a torus threaded by weak poloidal magnetic fields. 
In the case that the perturbations in the azimuthal direction 
increase, and non-axisymmetric structures form, any 3D aspects would
clearly appear in the spectra.

\subsection{Fitting to the flaring-state spectrum}
A striking fact is that 
{\it we cannot fit both the radio and X-ray data 
 simultaneously with the current MHD flow model}.
The MHD flow model can reproduce only a part of
the observations.  For example, 
it can fit the observed radio peak, if we assign
$\rho_0=1.6\times 10^{-14}\rm g \, cm^{-3}$,
but its flux largely exceeds the observed X-ray data
and the upper limits in the IR band.
This X-ray excess is caused by the strong free-free emission from 
the outer part of the flow ($r\gsim 10~r_{\rm s}$),
since the emissivity of free-free emission,
which is dominant in X-rays, is
$\varepsilon^{\rm ff} \propto \rho^2 T_{\rm e}^{1/2} \propto r^{1.5}$
for a density profile of $\rho \propto r$
and the $T_{\rm e}$ profile of $T_{\rm e} \propto r^{-1}$.
The entire luminosity is $\varepsilon^{\rm ff} d^3r \propto r^{3.5}dr$.

For comparison, we plotted the spectra for the emission
from only the inner region within $R=10~r_{\rm s}$ in Figure 1 
(see the upper dashed curve)
for the same density parameter,
$\rho_0=1.6\times 10^{-14}\rm g \, cm^{-3}$.
The electron temperature is 
$5.56 \times 10^9 {\rm K}$ (see Table \ref{Te}).
As shown in this figure, 
huge X-ray excesses disappear (cf. the case with $R=30~r_{\rm s}$) and
the model is successful in reproducing radio observations at around
$\log\nu\sim 11-12$ and is consistent with X-ray observations during
the flaring state (see the upper dashed curve).

The discrepancy between the upper thick-solid and 
upper dashed curves represents
the contribution from the outer parts ($r>10r_{\rm s}$).
This clearly demonstrates that
a huge X-ray excess is generated at the outer part of the flow
and we need to somehow remove these contributions to fit the date.

Alternatively, we can reduce the X-ray flux so as to fit the data
by employing a lower density parameter,
$\rho_0=5.6\times 10^{-16}\rm g \, cm^{-3}$
(see the lower thick-solid curve in Figure 1), but for such
a case, radio peak flux decreases in accordance with the reduced
density and, hence, is short of the radio luminosity.
We thus see that for this particular density profile,
it is totally impossible to reproduce the observations regardless of $\rho_0$
(see also Mineshige et al. 2002).

This situation resembles that of the CDAF (convection-dominated
accretion flow; see Ball, Narayan, \& QuataertBall 2001), 
since it has a flatter density profile like for the MHD flow (MMM01).
For $\rho \propto r^{-1/2}$ and $T \propto r^{-1}$, we have
$\varepsilon^{\rm ff} d^3r \propto r^{0.5}dr$,
indicating significant free-free emission from the outer parts.

Here, we stress that the emergent spectra starts to resemble
to the observed data if the magnetic field strengths 
are systematically under-estimated in the 3D MHD simulation,
that is, the actual magnetic fields are 
stronger than in the simulated data.
As was already mentioned in the previous subsection,
the radio peak luminosity and IR flux both increases
with an increase of the magnetic field strength.
Therefore, the emergent spectra are consistent with the 
observations if the actual magnetic fields are ten times 
stronger than the simulated data (see the thin solid curve).

Strictly speaking, the calculated spectrum also
deviates from the observed data at lower frequencies, $\log\nu<10$.
This inconsistency would be resolved 
if we consider synchrotron emission from nonthermal electrons
as was first claimed by Mahadevan (1998).
By a careful fitting to the up-to-dated spectrum,
Yuan, Quataert, \& Narayan (2003) concluded that
the low-frequency radio flux can be explained 
via the emission of nonthermal electrons,
which possess a small fraction of the electron thermal energy.

\subsection{Case of the quiescent spectrum}

What about the case of the quiescent spectrum?
Before attempting any fitting to the quiescent spectrum, 
we need to remark on the observational constraints of
the size of the emission region.
By the Chandra X-ray observations, Baganoff et al. (2001, 2003)
have claimed that the X-ray emission of Sgr A* is extended 
on scales of about 0.04 pc in the quiescent state.
However, Tan \& Draine (2004) suggested that about half of the
extended X-ray emission is due to dust scattering from an unresolved
source.  Therefore, we regard the observed X-ray data in the quiescent state 
to give an upper limit for the X-ray luminosity of the accretion flow
around the black hole.

To fit the X-ray flux in the quiescent state,
we need to adopt a small density parameter,
$\rho_0=2.0 \times 10^{-15}\rm g \, cm^{-3}$ (see the lower dashed curves),
but, then, the flow can not reproduce a radio peak.
In addition to this, the X-ray spectral slope also differs.
The X-ray excess decreases with a decrease in size of the
emission region, 
since the X-ray emission is dominated by the outer part,
as we already mentioned in the previous subsection.
We can adjust the X-ray luminosity to the observed value
in the quiescent state
by setting a much smaller emitting region, $R< 10~r_{\rm s}$.
However, the X-ray spectral slope
is still much steeper than the observed one.

To summarize, we
can reproduce the observed SED during the flaring 
state if the emitting region of the flow is restricted to be 
relatively small ($r<10~r_{\rm s}$), 
but cannot account for the quiescent-state spectrum 
for any choice of parameters.  
Note that a compact emission
region is needed to account for short-term variability, since
otherwise the variability timescale will be much longer in accordance
with the timescales in the outer zone (Mineshige et al. 2002).

\subsection{Time variation}
MHD simulations show that MHD flows are highly time-variable.
It is thus interesting to see how the resultant spectra change with
time and check if the calculated trend fits the observational ones.
We plot the time variations of the spectra in Figure 2.
\begin{figure}[h]
\epsscale{1.18}
\plotone{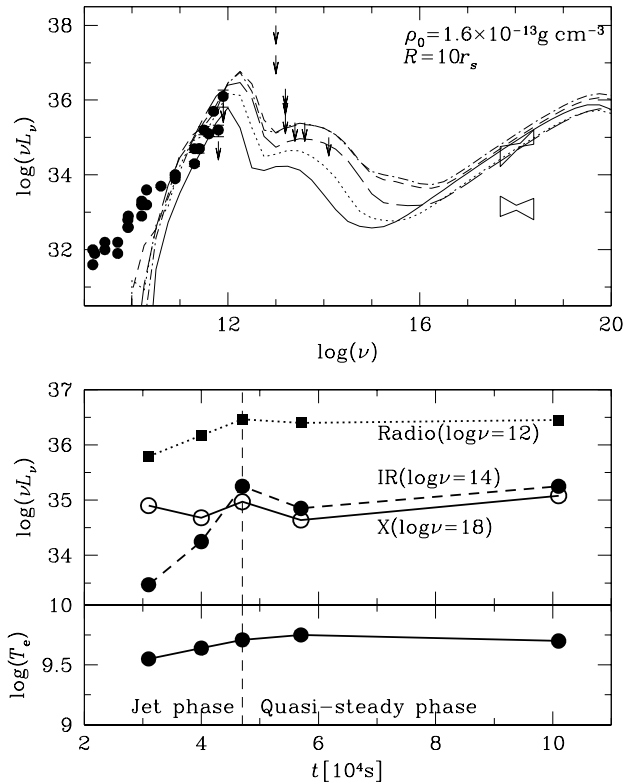}
\caption{
Spectral variations.
Upper panel: the resultant spectra 
at $t=3.1 \times 10^4$s (solid), $t=4.0 \times 10^4$s (dotted), 
$t=4.7 \times 10^4$s (dashed), $t=5.7 \times 10^4$s (dot-dashed), 
and $t=10.1 \times 10^4$s (long-dashed), respectively.
Here, we set $R=10~r_{\rm s}$ and $\rho_0=1.6\times 10^{-14} \rm g\,cm^{-3}$.
The lower panel:
time variations of radio, IR, and X-ray luminosities
as well as the electron temperature.
The electron temperature is nearly constant.
The radio and IR luminosities monotonously increase 
until the jet is produced ($t\leq 5\times 10^4$s) 
while X-rays fluctuate.
Radio luminosity is nearly constant at $t\geq 5\times 10^4$s,
but the flux slightly fluctuates at IR and X-ray wavelengths
and their amplitudes vary a few hundred percent. 
\label{fig2}
}
\end{figure}
Here, we again set $R=10~r_{\rm s}$ and 
$\rho_0=1.6\times 10^{-14} \rm g\,cm^{-3}$.
The calculated epochs are divided into two phases;
the jet phase ($t\lsim 5\times 10^4$s) 
and quasi-steady phase ($t\gsim 5\times 10^4$s).
The fluxes fluctuate in a wide range of frequencies
as shown in the upper panel,
although the electron temperature is kept nearly constant 
in the present study (see lower panel).

As shown in this figure, 
the radio and IR luminosity increases 
until the jet is generated ($t \lsim 5\times 10^4$s),
since magnetic fields are steadily being amplified in this phase.
In the quasi-steady phase ($t\gsim 5\times 10^4$s),
in contrast,
the radio luminosity is kept nearly constant,
whereas IR and X-ray luminosities fluctuate 
and their variation amplitudes amount to a few hundred percent. 
These results are roughly consistent with the observed trends
of multi-wavelength variability properties reported by
Ulrich, Maraschi, \& Urry (1997), who show that
the largest amplitude variations are in the soft X-rays
and that the optical-UV variation amplitude decreases systematically
with increasing wavelength in the LLAGNs.

Interestingly, the calculated flux at each wavelength does not always 
vary coherently.  We found that the X-ray luminosity decreases until
$t\sim 4\times 10^4$s, while the radio and IR luminosities increase.
Moreover, only the radio luminosity does not vary so much 
between $t=4.7\times 10^4$s and $t=5.7\times 10^4$s
in spite of the significant decrease in the IR and X-ray luminosities
(see lower panel).  A comparison of such trends with the observations
is left for future work.

The X-ray time-variation predicted by our results
is too small to account for the observed large-amplitude variations
(X-ray burst) reported by Baganoff et al. (2003), i.e., 
the X-ray luminosity in the flaring state is about 45 times larger
than that in the quiescent state.  Such an X-ray flaring event might
be produced by the nonthermal electrons, whereas we take only thermal
electrons into account in the present study.

\section{DISCUSSION: THE IMPLICATIONS OF OUR RESULTS}

We have calculated the emergent spectra based on the MHD simulation 
by KMS04 and demonstrated that the MHD flow model cannot account for
both of the radio and X-ray observations in the flaring state
simultaneously, unless we restrict the emission region to be compact.
We have also shown that it is problematic 
for reproducing the spectrum in the quiescent state.
In this section, we discuss what these results imply.

\subsection{flaring state}

We first focus our discussion on the case of the flaring state.
As was already mentioned in \S 3.2, 
the relatively compact emission region ($r<10~r_{\rm s}$) is required 
for the MHD flow model to reproduce the observed spectrum
in the flaring state of Sgr A*.
There are several possibilities which meet this requirement.

\subsubsection{Modifying density profiles}

A compact emission region would be realized, if the density profile
at large radii would be much steeper than that obtained by KMS04
or if the broad density peak found in KMS04 is somehow removed.
Since KMS04 provides only one specific example, it is necessary
to check the density profiles of other MHD simulations.
In Table \ref{MHD}
we summarize the density profiles of some representative simulations
with brief descriptions of their calculations.
\begin{table}[b]
\begin{center}
\caption{List of MHD models \label{MHD}}
\begin{tabular}{cccccc}
\tableline\tableline
ref. & $\Psi$ & I.C./B.C. & jet & $\rho$-profile & $\rho$-peak \\
\tableline
SP01 & {PN} & {$B_{\rm p}$-torus}
& {yes} & {$r^{0.1}(<10~r_{\rm s})$, $r^{-1}(>10~r_{\rm s})$} & {$10~r_{\rm s}$} \\ 
MMM01 & {N} & {$B_\phi$-torus}
& {no} & {$r^{-1/2}$} & {---} \\ 
HB02 & {PN} & {$B_{\rm p}$-torus}
& {yes} & {$r^0$+torus($r<10~r_{\rm s}$)} & {$ 5~r_{\rm s}$} \\ 
INA03 & {PN} & {injection}
& {yes} & {$r^{-1/2}(<10~r_{\rm s})$, $r^{-1}(>10~r_{\rm s})$ } & {$ 3~r_{\rm s}$} \\ 
MM03 & {PN} & {$B_\phi$-torus}
& {no} & {$r^0$} & {$ 3~r_{\rm s}$} \\ 
PB03 & {PN} & {Bondi flow}
& {yes} & {$r^{0.2}(<7~r_{\rm s})$, $r^{-1}(>7~r_{\rm s})$ } & {$ 7~r_{\rm s}$} \\ 
KMS04 & {PN} & {$B_{\rm p}$-torus}
& {yes} & {$r^{1}(<20~r_{\rm s})$, $r^{-1}(>20~r_{\rm s})$} & {$ 20~r_{\rm s}$} \\ 
Our 2D & {PN} & {$B_{\rm p}$-torus}
& {yes} & {$r^{1}(<7~r_{\rm s})$, $r^{-1}(>7~r_{\rm s})$} & {$ 7~r_{\rm s}$} \\ 
\tableline
\end{tabular}
\end{center}
\tablenotetext{}{
SP01: Stone \& Pringle (2001);
MMM01: Machida, Matsumoto \& Mineshige (2001);
HB02: Hawley \& Balbus (2002);
INA03: Igumenshchev, Narayan, \& Abramowicz (2003);
MM03: Machida \& Matsumoto (2003);
Our 2D: our 2D MHD simulations;
$\Psi$: gravitational potential;
N: Newtonian; PN: pseudo-Newtonian;
I.C./B.C.: initial or boundary condition;
$B_{\phi}$- or $B_{\rm p}$-torus:
A torus threaded by weak localized toroidal or poloidal magnetic fields
is initially assumed.;
injection: Magnetized matter is continuously injected 
into the computational domain near the outer boundary
}
\end{table}

SP01, 
HB02, and KMS04
adopted similar initial conditions; namely,
they started calculations with a torus 
threaded by weak localized poloidal magnetic fields
and assumed no further mass input.  
They all investigated the evolution of the torus under
the pseudo-Newtonian potential.  
INA03 and Proga \& Begelman (2003: hereafter refereed to as PB03)
set a different situation; INA03
continuously injected magnetized matter 
into the computational domain from near the outer boundary,
and PB03 started the simulations with Bondi accretion flow,
whose angular momentum is zero except for the outer part of the flow.
Note that 
SP01 and PB03 performed 2D simulations, while others did 3D simulations.
Despite some differences in simulations,
both of SP01, INA03, and PB03 obtained a relatively steep density
profile, $\rho \propto r^{-1}$.  No obvious density peak is found in
their simulations.

HB02 obtained a similar density profile but with a remarkable peak
at $\sim 5~r_{\rm s}$, which may be a remnant of the initial torus.
Caution should be taken here, since this peak may be transient.
Similarly, the density peak found at $\sim 20~r_{\rm s}$ in KMS04,
which mainly contributes to X-ray emissions,
might be a direct consequence of the initial torus
located at $40~r_{\rm s}$ at the beginning of the simulations.
If this density peak is removed by long term numerical calculations,
we expect that the large X-ray excess may disappear and 
the resultant spectra can be consistent with the observations,
even if we do not artificially restrict the emission region.

To see if this is the case, we made the following test.
We performed 2D MHD simulations, 
starting from the similar initial condition to 
that of KMS04 but with the initial torus being put 
a much larger radius, around $\gsim 100 ~r_{\rm s}$.
After a long time interval, we have confirmed that
the broad density peak, a remnant of the initial torus, disappears.
The resultant density structure in this 2D model (at $\tilde{t}=8200$)
is plotted by the solid curve in the upper panel 
in Figure 3 with that of KMS04.
The density profile of our 2D model is roughly proportional to $r^{-1}$
at $r>7~r_{\rm s}$ and roughly agrees with the simulations by SP01, INA03
and PB03.
\begin{figure}[h]
\epsscale{1.18}
\plotone{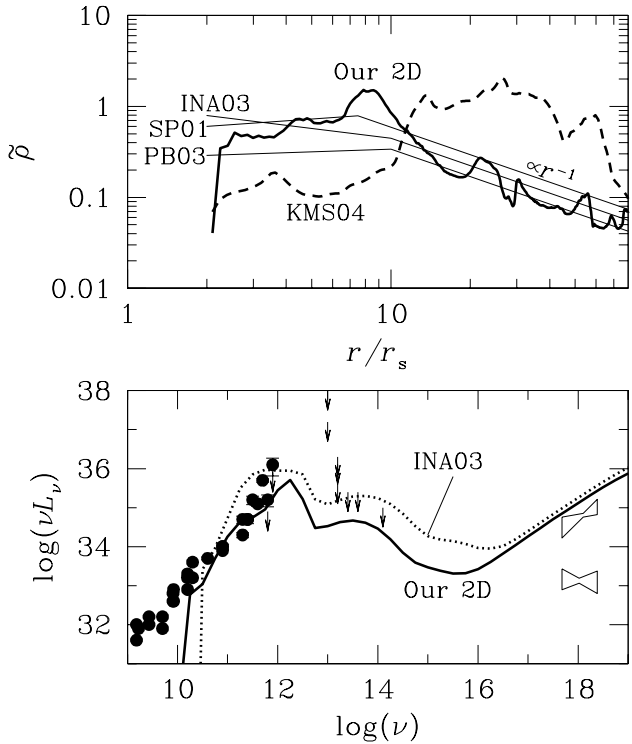}
\caption{
Upper panel:
The normalized density profile given by our 2D MHD simulations (thick solid)
and KMS04 (dashed).
The density peak in KMS04 is around $20~r_{\rm s}$,
in contrast, it is near the black hole ($r\sim 7~r_{\rm s}$)
in the case of our 2D simulations.
The thin solid curves indicate the schematic density profile 
given by INA03 (3D simulations), SP01, and PB03 (2D simulations).
They showed a similar density profile, $\rho \propto r^{-1}$,
with our 2D MHD simulations at the outer part ($r>10~r_{\rm s}$).
Lower panel:
The solid curve is the spectra of our 2D MHD model.
We consider the emission from the extended region of $r<30~r_{\rm s}$.
The normalization of density is set to be 
$\rho_0=3.0\times 10^{-15}{\rm g \, cm^{-3}}$,
and the calculated electron temperature is shown in Table \ref{Te}.
The dotted curve represents the spectra of the INA03 model.
Here, we use the spherical flow model, 
which is constructed based on the 3D MHD simulations of INA03.
In this model, the density and proton temperature profiles 
are $\rho = 7.0\times 10^{-15} (r/r_{\rm s})^{-1} \rm g\, cm^{-3}$ and
$T_{\rm p} = 8.6\times 10^{11} (r/r_{\rm s})^{-1} \rm K$, respectively.
We set $T_{\rm e}=5\times 10^9\rm K$ and plasma-$\beta=3.0$.
Our 2D MHD model as well as the INA03 model can reproduce 
the observed data during the flaring state.
\label{fig3}
}
\end{figure}

We, next, calculate the spectrum based on our 2D model 
and plot it in Figure 3 (lower panel).
The adopted density parameter is 
$\rho_0=3.0 \times 10^{-15} \rm g \, cm^{-3}$,
and the size of the emitting region is $R=30~r_{\rm s}$.
The calculated electron temperature is shown in Table \ref{Te}.
Clearly, our 2D MHD model can nicely fit the observed data points
during the flaring state.  
Thus, it is very likely that if we continue MHD simulations from the
data by KMS04 further, we may be able to fit the observational data.
We need caution, however, because behavior of the magnetic fields in
2D and 3D simulations may be distinct.  We cannot trust the 2D data so
much.  We need to perform longer 3D simulations to confirm our
tentative conclusions.

It is expected that the MHD models by INA03, SP01, and PB03
can also reproduce the observed SED during the flaring state, since 
these models have similar density profiles to those of our 2D model
(see the upper panel of Figure 3).
We demonstrate it by calculating the spectrum
from the regions at $r<30~r_{\rm s}$ of the spherical flow model,
which is constructed based on the 3D MHD simulations of INA03
(see the dotted curve in the lower panel of Figure 3).
In this model the density and proton temperature profiles 
are set to be $\rho = 7.0\times 10^{-15} (r/r_{\rm s})^{-1} \rm g\, cm^{-3}$ 
and $T_{\rm p} = 8.6\times 10^{11} (r/r_{\rm s})^{-1} \rm K$, respectively.
The electron temperature and the plasma-$\beta$ are 
assumed to be constant in the calculating box,
$T_{\rm e}=5\times 10^9\rm K$ and $\beta=3$.
The INA03 model shows a similar SED with that of our 2D model,
and is consistent with the observed data
in the flaring state.
The other MHD models by HB02, MMM01, and MM03, on the other hand,
have flatter density profiles, $\rho\propto r^{-a}$ with $a=0-1/2$
(see Table \ref{MHD}).  Hence, 
they will produce an X-ray excess as was in the case of our simulation.

\subsubsection{Moderate radiative cooling}

A distinct way of possibly producing a steeper density profile is 
to introduce moderate radiative cooling (Mineshige et al. 2002).
By the terminology of RIAF it means that radiative loss is inefficient.
However, it may happen that
radiative cooling is only partly and transiently important.
If we set the cooling timescale to be comparable to the accretion timescale, 
we expect that only dense parts can efficiently be cooled. 
Since the density is generally higher in the innermost region, 
cooling results in further mass concentration towards the center,
a preferable condition to fit the observations.  
However, this is only a possibility and
we need 3D MHD simulations to confirm this idea.

Another possibility is that the outer portions of the disk
may become radiation efficient flow. 
Even if the density profile does not change,
the emission from the outer parts can be reduced,
if the electron temperature there were
much lower than that of the inner part.
In KMS04 and most of other MHD simulations, 
the one-temperature plasma is assumed 
and the cooling of protons is neglected.
However, the proton thermal energy may be converted 
into the electron energy due to the effective Coulomb collisions 
and be radiated away in the outer zones.
Further study requires
3D MHD simulations coupled with the electron energy equation
and radiative cooling.

\subsection{Quiescent state}

The case of the quiescent spectrum will be much more difficult to
reproduce by the MHD model. We have seen that 
the calculated spectral slope was much steeper than 
that of the X-ray observations in the quiescent state of Sgr A*.
We can reduce the X-ray flux but cannot easily change the spectral slope 
in the X-ray range.
We find two possibilities to make a flatter spectral slope.

\subsubsection{Nonthermal electrons}

This difficulty may be removed, if emission from the nonthermal 
electrons dominates over that from thermal ones.
By considering emission from nonthermal electrons,
Yuan, Quataert, \& Narayan (2003) nicely reproduced
the observed spectrum in the quiescent state of Sgr A*.
[However, they prescribe an arbitrary energy distribution of the
nonthermal electrons with more free parameters
in order to give a good fit to the observations.]
This is a very attractive possibility, but 
it is hard to prescribe an energy distribution of nonthermal electrons
because of a lack of good theory of particle acceleration.  Thus we
need to introduce parameters describing the electron energy
distribution and the results strongly depend on these parameters.

\subsubsection{Compton scattering in disk coronae}

The spectral slope at the X-ray band would be flatter if inverse
Compton scattering is more enhanced, compared with other processes, 
and becomes dominant in the X-ray range.
Note that the calculated steep spectral slopes shown in
Figure 1 are due to bremsstrahlung emissions from the outer parts.
The relative importance of the Comptonization inside the corona is 
proportional to the column density, $\sim \rho H$
(with $H$ being the half-thickness of the corona),
while the bremsstrahlung emissivity is proportional to $\rho^2 H$.
Thus, the presence of an extended, tenuous corona 
may resolve the problem.  However,
the formation mechanism of the corona is still an open issue
and it is not easy to probe this possibility

\section{CONCLUSIONS}

In order to test the 3D MHD flow simulation
as a model of optically-thin, high-temperature accretion flows
through comparison with observations,
we studied the spectral properties of magnetized accretion flows 
based on the 3D MHD simulation data.
Surprisingly, such comparative studies have not been
well investigated so far in spite of a large number of MHD
simulations having been performed recently.
We summarize our results as followings:

\begin{enumerate}
\item 
We found that the MHD model cannot reproduce the observed SED 
in the flaring state of Sgr A* without substantial modification. 
If we use the data by KMS04, we need to require that the emission 
region is restricted to be compact ($r<10~r_{\rm s}$), but this could be 
due to the particular initial condition. Some other MHD 
models with a steeper density profile ($\rho \propto r^{-1}$)
and without a broad density peak ($r>10r_{\rm s}$) can fit the 
observations, if the emissions from regions of 
$r\lsim 30r_{\rm s}$ mainly contributes to the SEDs.

\item
The MHD flow can not generally 
reproduce the spectrum (spectral shape, in particular)
in the quiescent state.
Significant contributions by nonthermal electrons should resolve
this issue.  It is also possible that observational X-ray data 
may contain emissions from regions other than the vicinity of 
the black hole.  

\item
The MHD flow predicts substantial and incoherent time variations 
in the emergent spectrum.  The predicted variation amplitude is,
however, too small to account for the large-amplitude X-ray burst.
\end{enumerate}

\acknowledgments
The authors would like to thank the anonymous referee 
for important comments and suggestions. 
The calculations were carried out at 
Yukawa Institute for Theoretical Physics at
Kyoto University 
and the Department of Physics at Rikkyo University.
This work is 
supported in part by Research Fellowship of the Japan Society
for the Promotion of Science for Young Scientists, 02796 (KO)
and the Grants-in-Aid of the
Ministry of Education, Science, Culture, and Sport, 
(14079205, 16340057)
and by a Grant-in-Aid for the 21st Century COE
 {\lq\lq}Center for Diversity and Universality in Physics" (SM).

\end{document}